\documentclass[preprint, amsmath,amssymb,aps]{revtex4-1}

\usepackage{graphicx}% Include figure files
\usepackage{dcolumn}% Align table columns on decimal point
\usepackage{bm}% bold math
\usepackage{natbib}
\providecommand{\keywords}[1]{\textbf{\textit{Index terms---}} #1}
\usepackage[mathlines]{lineno}%
\usepackage[colorinlistoftodos]{todonotes}
\usepackage[colorlinks=true, allcolors=blue]{hyperref}
\begin{document}

\preprint{PRB}

\title{Metal-Insulator transition in $8-Pmmn$ Borophene under perpendicular incidence of electromagnetic radiation}% Force line breaks with \\

\author{Abdiel E. Champo}
\affiliation{
Facultad de Ciencias en F\'isica y Matem\'aticas, Universidad Aut\'onoma de Chiapas, Carretera Emiliano Zapata, Km. 8, Rancho San Francisco, 29050, Tuxtla Guti\'errez, Chiapas, M\'exico.
}

\author{Gerardo G. Naumis}
\email{naumis@fisica.unam.mx}
\affiliation{Departamento de Sistemas Complejos, Instituto de Fisica, Universidad Nacional Aut\'onoma de M\'exico, Apartado Postal 20-364,01000,Ciudad de M\'exico, M\'exico.
}

\date{\today}

\begin{abstract}

The energy spectrum for the problem of $8-Pmmn$ borophene's electronic carriers under perpendicular incidence of electromagnetic waves is studied without the use of any perturbative technique. This allows to study the effects of very strong fields. To obtain the spectrum and wavefunctions, the time-dependent Dirac equation is solved by using a frame moving with the space-time cone of the wave, i.e., by transforming the equation into an ordinary differential equation in terms of the wave-phase, leading to an electron-wave quasiparticle. The limiting case of strong fields is thus analyzed.The resulting eigenfunctions obey a generalized Mathieu equation,i.e., of a classical parametric pendulum.  The energy spectrum presents bands, and a gap at the Fermi energy. The gaps are due to the space-time diffraction of electrons in phase with the electromagnetic field, i.e., electrons in borophene acquire an effective mass under strong electromagnetic radiation
\begin{description}
\item[Keywords]
borophene; electromagnetic waves; Mathieu equation; gaps 
\end{description}
\end{abstract}

%Use showkeys class option if keyword
\maketitle

%\tableofcontents

\section{Introduction}
In recent times, Dirac materials have attracted intense research interest after the most celebrated discovery of an atomically two- dimensional (2D) hexagonal carbon allotrope, graphene\cite{Novoselov666}, owing to their peculiar band structure and applications in next generation of nanoelectronics \cite{RevModPhys.81.109,RevModPhys.83.407,0034-4885-80-9-096501}.

Following the seminal discovery of graphene, great efforts have been paid to search for new Dirac materials which can host massless Dirac fermions (MDF) \cite{doi:10.1093/nsr/nwu080, doi:10.1080/00018732.2014.927109}, especially in monolayer structures.

\begin{figure} 
\includegraphics[scale=0.20]{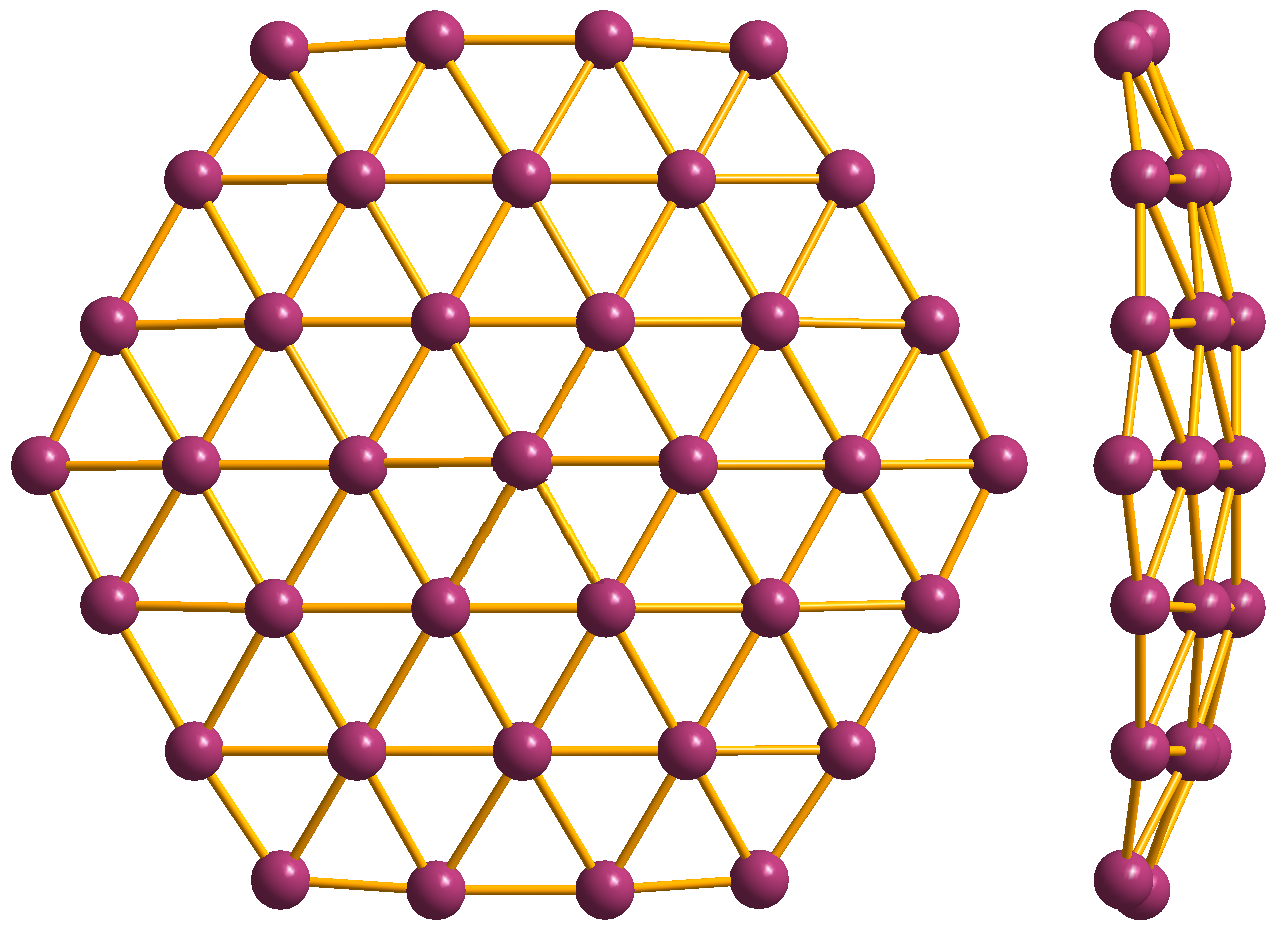}
\caption{\label{Borophene lattice} Borophene lattice for a phase with space group $8-Pmmn$.}
\end{figure}

Recently, there has been intense research interest in the synthesis of 2D crystalline boron structures, referred to as borophenes. See as an example  the structure shown in Fig. \ref{Borophene lattice}.  Boron is a fascinating element due to its chemical and structural complexity, and boron- based nanomaterial of various dimensions have attracted a lot of attention \cite{C7CS00261K} . For example, two-dimensional phases of boron with space groups $Pmmm$ and $Pmmn$ and hosting MDF were theoretically predicted \cite{PhysRevLett.112.085502}. Several attempts have been made to synthesize a stable structure of borophene, but only three different quasi-2D structures of borophene have been synthesized \cite{Mannix1513}.  Various numerical experiments have predicted a large number of borophene structures with various geometries and symmetries \cite{C6CP05405F,PhysRevLett.112.085502}. The orthorhombic $8-Pmmn$ borophene is one of the energetically stable structures, having ground state energy lower that of the $\alpha-$sheet structures and its analogues.
The $Pmmn$ boron structures have two nonequivalent sublattices. The coupling and buckling between two sublattices and vacancy give rise to energetic stability as well as tilted anisotropic Dirac cones \cite{PhysRevB.93.241405}. The coupling between different sublattices enhances the strength of the boron-boron bonds and hence gives rise to structural stability. The finite thickness is required for energetic stability of 2D boron allotropes. The orthorombic $8-Pmmn$ borophene possesses tilted anisotropic Dirac cones and is a zero-gap semiconductor. It can be thought of as topologically equivalent to the distorted graphene.

A tight binding-model of $8-Pmmn$ borophene has been recently developed \cite{PhysRevB.97.125424,PhysRevB.94.165403} and an effective low-energy Hamiltonian in the vicinity of Dirac points was proposed on symmetry consideration. Pseudomagnetic  fields were also predicted similar to those in strained graphene \cite{0034-4885-80-9-096501,PhysRevLett.103.046801,0953-8984-28-2-025301,OLIVALEYVA20152645,PhysRevB.88.085430} 
and its relationship with electronic \cite{PhysRevB.92.035406,GarciaNaumis2017}  and optical conductivity \cite{PhysRevB.93.035439,PhysRevB.89.241404}.
 
 In $8-Pmmn$ borophene, the effective low-energy Hamiltonian was used to study the plasmon dispersion and screening properties by calculating the density- density response function \cite{PhysRevB.96.035410}, the optical conductivity \cite{PhysRevB.96.155418},  Weiss oscillations\cite{PhysRevB.96.235405} and oblique Klein tunneling \cite{PhysRevB.97.235440}. The fast growing experimental confirmation of various borophene monolayers make $8-Pmmn$ borophene promising. However, the calculated properties of electrons under electromagnetic fields are all based on perturbative approaches. Yet,  it is known that in graphene interesting non-linear effects appear when strong fields are applied \cite{PhysRevB.78.201406,doi:10.1080/14786431003757794}.

In this work, we solve the problem of borophene's electron behaviour in the presence of a strong electromagnetic plane wave. As a result, we are able to find the spectrum, wavefunctions and a dynamic gap opening. This is like if electron in borophene acquire an effective mass under electromagnetic radiation. It is important to remark that strain affects the optoelectronics properties of 2D materials, such as phosphorene \cite{Mehboudi5888} or graphene \cite{0034-4885-80-9-096501}, and these effects can be also studied by using the present methodology, as strain and flexural waves can be considered as pseudoelectromagnetic waves \cite{0953-8984-28-2-025301,doi:10.1002/pssr.201800072}.

The paper is organized as follows. In Sec. \ref{Model Hamiltonian}, we introduce the low-energy effective Hamiltonian and obtain the Hill's equation that solve the problem of borophene's electron behaviour under electromagnetic radiation without the need of any approximation. Section \ref{Spectre and Eigenfunctions} is devoted to solve the Mathieu's equation with the strong electromagnetic field approximation or long wavelength and we obtain as solution the Mathieu cosine and Mathieu sine functions; beside, we study the stability chart of the solutions. Finally, we summarize and conclude in Sec. \ref{Conclusions}.

\section{Model Hamiltonian \label{Model Hamiltonian}} 
\begin{figure} 
\includegraphics[scale=0.4]{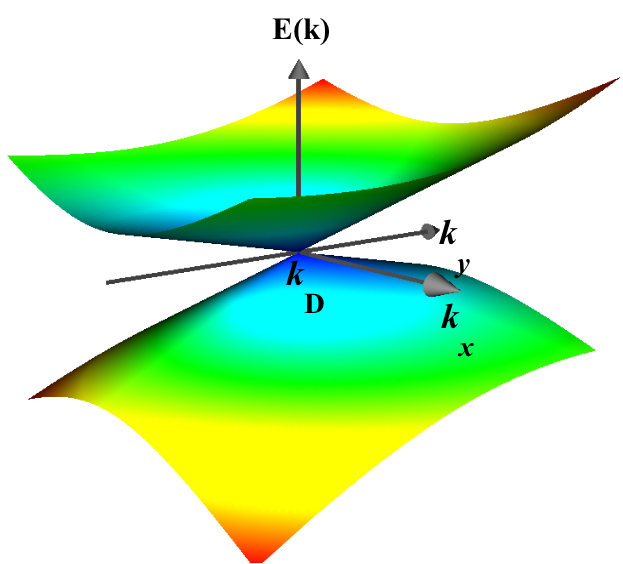}
\caption{\label{Borophene spectre} Plot of the energy dispersion $E(\textbf{k})$ as a function of $\textbf{k}$ (see Eq. \eqref{Energy of borophene isolated}), in the region around a Dirac cone. Notice the tilted anisotropy of the Dirac Cone.}
\end{figure}

We start with the single-particle low-energy effective model Hamiltonian  of the tilted anisotropic Dirac cones as \cite{PhysRevB.94.165403},
\begin{equation} \label{Hamiltonian  Pristine Borophene}
\hat{H}= \varrho (v_{x} \hat{P}_{x} \sigma_{x} + v_{y} \hat{P}_{y} \sigma_{y} +v_{t} \hat{P}_{y} \sigma_{0})
\end{equation}
where the first two terms correspond to the kinetic energy term and the last term described the tilted nature of Dirac cones. The two Dirac points ${\bf{k}}= \pm {\bf{k_{D}}}$ are described by the valley index $\varrho= \pm 1$. The three velocities along each coordinate  are given by $\{v_{x},v_{y},v_{t}\}= \{ 0{.}86,0{.}69,0{.}32\}$ in units of $v_{F}= 10^{6} \,\,\, m/s$. The velocity $v_{t}$ arises due to the tilting of the Dirac cones. Also, $(\sigma_{x},\sigma_{y})$ are the Pauli matrices and $\sigma_{0}$ is the identity matrix. $\hat{P}_{x}, \hat{P}_{y}$ are the  electron momentum operators. The energy dispersion of the above Hamiltonian can be readily obtained as \cite{PhysRevB.96.235405},
\begin{equation} \label{Energy of borophene isolated}
E_{\lambda,k}^{\varrho} = \varrho \hbar v_{t} k_{y}+ \lambda \hbar \sqrt{v_{x}^{2}k_{x}^{2}+v_{y}^{2}k_{y}^{2}}
\end{equation}
and 
\begin{equation}
\psi_{\lambda,\bf{k}}^{\varrho}= \varrho \frac{e^{i \bf{k \cdot r}}}{\sqrt{2}} \left( \begin{array}{lcc}
1 \\
\lambda e^{i \Theta} 
\end{array} \right)
\end{equation}
where $\lambda= \pm 1$ is the band index, $\Theta= \tan ^{-1} (v_{y}k_{y}/v_{x}k_{x})$ and the $2D$ momentum vector is given by $\boldsymbol{k}=(k_{x},k_{y})$. 
The energy dispersion for the $K$ valley is shown in Fig. \ref{Borophene spectre}.

\subsection{Inclusion of electromagnetic field}
For borophene irradiated by a propagating electromagnetic field perpendicular to the graphene's plane,  the Dirac Hamiltonian can be obtained by using the minimal coupling, 
\begin{equation} \label{Hamiltonian borophene under electromagnetic field} 
\hat{H}= \left( \begin{array}{lcc}
v_{t} \hat{\Pi}_{y} & v_{x} \hat{\Pi}_{x}- i v_{y} \hat{\Pi}_{y} \\
v_{x} \hat{\Pi}_{x}+ i v_{y} \hat{\Pi}_{y} & v_{t} \hat{\Pi}_{y}
\end{array} \right)
\end{equation}
where, $\hat{\Pi}= \hat{P}-(e/c) {\bf{A}}$ and $\bf{A}$ is the vector potential of the applied electromagnetic field, given by ${\bf{A}}= \frac{E_{0}}{\Omega} \cos(Gz-\Omega t) (\cos(\theta),\sin(\theta))$, where $\Omega$ is the frequency of the wave and $E_{0}$ is the amplitude of the electric field, taken as a constant. The dynamics of charge carriers in graphene is governed by a time-dependent Dirac equation:
\begin{equation} \label{Dirac equation}
\hat{H}(x,y,t){\bf{\Psi}}(x,y,t)= i \hbar \frac{\partial}{\partial t} {\bf{\Psi}}(x,y,t)
\end{equation}
where,
\begin{equation} \label{Spinor Psi}
{\bf{\Psi}}(x,y,t)= \left( \begin{array}{lcc}
\Psi_{A}(x,y,t) \\
\Psi_{B}(x,y,t)
\end{array} \right)
\end{equation}
is a two component spinor which give the electron's wavefunction on each borophene sublattice, denoted by $A$ and $B$. To find the eigenstates and eigenenergies we use instead Pauli matrices and spinors. First we write the equations of motion for each component of the spinor,

\begin{equation} \label{Dirac Equation PsiA}
\begin{split}
 i \hbar \frac{\partial \Psi_{A}(x,y,t)}{\partial t}&= \left( v_{x} \hat{\Pi}_{x} - i v_{y} \hat{\Pi}_{y} \right) \Psi_{B} (x,y,t) \\
 &+v_{t} \hat{\Pi}_{y} \Psi_{A}(x,y,t) 	
\end{split}
\end{equation}

\begin{equation} \label{Dirac Equation PsiB}
\begin{split}
i \hbar \frac{\partial \Psi_{B}(x,y,t)}{\partial t} &=
 \left( v_{x} \hat{\Pi}_{x} - i v_{y} \hat{\Pi}_{y} \right) \Psi_{A} (x,y,t) \\
 & +v_{t} \hat{\Pi}_{y} \Psi_{B}(x,y,t)
\end{split}
\end{equation}
The most important step in the solution of this problem is to propose a solution of the form
\begin{equation} \label{Form of a solution}
\Psi_{\rho}(x,y,t)=e^{i({\bf{k}}\cdot {\bf{r}}-Et/\hbar)} \Phi_{\rho}( \phi)
\end{equation}
where we have defined the phase $\phi$ of the electromagnetic wave as $\phi=Gz- \Omega t$, and $\Phi_{\rho}(\phi)$ is a function to be determined for $\rho=A,B$. This ansatz is equivalent to consider the problem in the space-time frame of the moving wave. As detailed in the Appendix, the system of differential equations can be further rewritten in terms of two new functions $\Gamma_{A}(\phi)$ and $\Gamma_{B}(\phi)$, defined by:
\begin{equation} \label{transformation PhiRho}
\Gamma_{\rho} (\phi)=\exp{\lbrace-\frac{i}{c}[(v_{t}\tilde{k}_{y}-v_{y} \tilde{E})\phi-v_{t} \tilde{\xi} \sin \phi \sin \theta]\rbrace}\Phi_{\rho}(\phi)   
\end{equation}
and obtain
\begin{equation} \label{Equation GammaA}
\frac{d \Gamma_{A}(\phi)}{d \phi}= iC^{*}(\phi) \Gamma_{B}(\phi)
\end{equation}
\begin{equation} \label{Equation GammaB}
\frac{d \Gamma_{B}(\phi)}{d \phi}= i C(\phi) \Gamma_{A}(\phi)
\end{equation}

where 
\begin{equation}
 C(\phi)=\frac{[v_{x}(\tilde{k}_{x}-\tilde{\xi}\cos \phi \cos \theta)+iv_{y}(\tilde{k}_{y}-\tilde{\xi}\cos \phi \sin \theta)]}{c}
\end{equation}

and the other are adimensional variables, defined as $\tilde{E}= E/(\hbar v_{y} G)$, $\tilde{{\bf{k}}}=  {\bf{k}}/G$, $\tilde{\xi}=eE_{o}/(c \hbar G \Omega)$. Finally, $c= \Omega/G$ is the light velocity.

As explained in the Appendix, Equations \eqref{Equation GammaA} and \eqref{Equation GammaB} can be written as a single second order ordinary differential equation. In the resulting equation, we consider the transformation:
\begin{equation} \label{Second transformation GammaA}
\Gamma_{A}(\phi)= e^{-i (\beta/2)} \chi_{A} (\phi)
\end{equation}
\begin{equation} \label{Second transformation GammaB}
\Gamma_{B}(\phi)= e^{i (\beta/2)}  \chi_{B} (\phi)
\end{equation}
where $\beta= \arctan((v_{y}/v_{x})\tan \theta)$; besides, if we consider that:
\begin{equation} \label{Spinor Chi}
{\bf{\chi}}(\phi)= \left(\begin{array}{lcc}
\chi_{A}(\phi) \\
\chi_{B} (\phi)
\end{array} \right)
\end{equation}
we finally obtain that the $\chi(\phi)$ functions follow a Hill's equation,
\begin{equation} \label{Second Order Equation Spinor Chi 1}
\frac{d^{2}}{d \phi^{2}} {\bf{\chi}}(\phi)+ F(\phi){\bf{\chi}}(\phi)=0
\end{equation}
with $F(\phi)$ defined as:
\begin{eqnarray} \label{Definition of F function}
F(\phi)= \left( \frac{1}{\hbar \Omega} \right)^{2} \left[  \zeta^{2} \cos^{2} \phi - 2 \zeta \frac{\vec{\nu} \cdot \vec{\kappa}}{|\vec{\nu}|} \cos \phi \right] \nonumber \\ 
+\left( \frac{\epsilon}{\hbar \Omega} \right)^{2}- i\frac{\zeta}{\hbar \Omega}  \sigma_{x} \sin \phi  
\end{eqnarray}
where $\epsilon= \hbar \sqrt{(v_{x} k_{x})^{2} +(v_{y}k_{y})^{2}}$, \,\, $\vec{\kappa}= \hbar (v_{x} k_{x}, v_{y} k_{y})$,\\
$\zeta= (eE_{0}/c \Omega) \sqrt{v_{x}^{2}\cos^{2} \theta + v_{y}^{2} \sin^{2} \theta}$  \,\,\, and
 \\ $\vec{\nu}=(v_{x} \cos \theta, v_{y} \sin \theta)$.

\section{Spectrum and Eigenfunctions \label{Spectre and Eigenfunctions}} 
The resulting Hill equation, given by expression Eq. ( \ref{Second Order Equation Spinor Chi 1}), is difficult to be solved analytically for all cases. 
Yet, the most interesting case for the physics of the problem is the limit of intense  applied electric fields or long wavelengths ($ E_{0}/\hbar \Omega^{2} >> 300)$, since  other limits can be tackled using perturbative approachs.

%\subsection{Intense electric fields or long wavelengths ($\bf{e E_{0}/\hbar \Omega^{2} >> 300)}$}
For this particular case $\zeta/ \hbar \Omega >>1$, and thus in Equation \eqref{Second Order Equation Spinor Chi 1} we can neglect linear terms in $\zeta/ \hbar \Omega$. Also, in what follows we take $\vec{\nu} \cdot \vec{\kappa}=0$, which is basically an initial condition, in order to simplify the equations, although the general case can be solved in a similar way. The following equation is obtained for $\chi( \phi)$:
\begin{equation} \label{Second Order Equation with intense electric field}
\chi''(\phi)+ \left \lbrace \frac{\epsilon^{2}}{(\hbar \Omega)^{2}}+ \left(\frac{\zeta \cos \phi}{\hbar \Omega} \right)^{2} \right \rbrace \chi (\phi)=0
\end{equation}
\begin{figure} [h!]
\includegraphics[width=9cm,height=11cm]{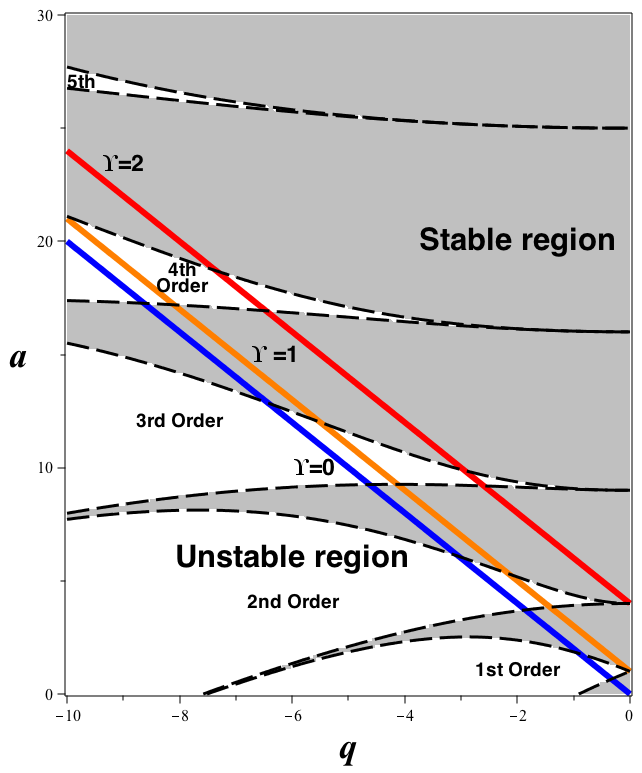}
\caption{\label{Spectrum of Mathieu equations} Stability chart as a function of the non-dimensional parameters $a$ and $q$ for Mathieu solutions. The regions of stability (gray domains) and instability (white domains) are divided by the characteristic curves $a_{n}(q)$ and $b_{n}(q)$ (dashed lines). This spectrum is for the case of intense electric fields or long wavelengths. In this case there is an constraint due to Eq. \eqref{definition of a}. For fixed energies $\epsilon$, this condition is equivalent to draw a set of parallel lines as shown in the figure, where $\Upsilon= \epsilon/ \hbar \Omega$. The line $a=-2q$, i.e., $\Upsilon=0$, divides the spectrum into two zones, the zone on the left side of this line are strictly not allowed, while the values on right side are allowed. A simple analysis of this figure reveals bands separated by energy gaps. The order of the gaps is indicated.}
\end{figure}
Using the relation $\cos^{2}(\phi)=\frac{1}{2} (1+cos(2 \phi))$ we can write,
\begin{equation}
\chi''(\phi)+\left \lbrace \left( \frac{\epsilon}{\hbar \Omega}\right)^{2} + \left( \frac{\zeta}{\hbar \Omega} \right)^{2} \left[ \frac{1}{2} (1+\cos(2 \phi)) \right] \right\rbrace \chi(\phi)=0
\end{equation}
and defining, 
\begin{equation} \label{definition of q}
q= - \left( \frac{\zeta}{2 \hbar \Omega} \right)^{2}
\end{equation}
and, 
\begin{equation} \label{definition of a}
a= \left(\frac{\epsilon}{\hbar \Omega}\right)^{2}-2q
\end{equation}
Thus, Equation \eqref{Second Order Equation with intense electric field} is  transformed into the following Mathieu equation,
\begin{equation} \label{Mathieu equation 1}
\frac{d^{2}}{d \phi^{2}}\chi(\phi)+ \left[a -2q \cos(2 \phi) \right] \chi(\phi)=0
\end{equation}
As is well known, this Mathieu equation describes a parametric pendulum, in which there is an interplay between two frequencies, one is the fundamental of the pendulum, determined by $\sqrt{a}$, and the other is the frequency of the cosine driving. The parameter $q$ measures the coupling between the natural and driving frequencies leading to an interesting resonance phase-diagram.

In our problem, the general solutions to the components $\chi_{A}$ y $\chi_{B}$ are linear combinations of the Mathieu cosine $\mathcal{C}(a,q,\phi)$ and Mathieu sine $\mathcal{S}(a,q,\phi)$ functions. Nevertheless, taking into account that when the electromagnetic field is switched out the wavefunction  $\bf{\Psi}$ must reduce to a free- particle wavefunction, we obtain that,
\begin{equation}
\begin{split}
\boldsymbol{\Psi}(x,y,t)&= N \,\,e^{i(\boldsymbol{k}\cdot \boldsymbol{r}- Et/\hbar- \beta/2)} \,\, e^{\frac{i}{c} \left[(v_{t}\tilde{k}_{y}-v_{y}\tilde{E})\phi-v_{t}\tilde{\xi}\sin \phi \sin \theta \right]} \\
& \times \left(\mathcal{C}(a,q,\phi)+ i\mathcal{S}(a,q,\phi)\right) \left( \begin{array}{lcc}
1 \\
\lambda e^{i(\beta+ \Theta)}
\end{array} \right)
\end{split}
\end{equation}
where $\beta=\tan^{-1}\left[\left(v_{y}/v_{x} \right) \tan \theta \right]$, $\Theta= \tan^{-1} \left( v_{y} k_{y}/ v_{x} k_{x} \right)$, $N$ is a normalization constant and $\lambda= \pm 1$ denotes the conduction and valence bands, respectively

\begin{figure}[h!]\label{Gap}
 \includegraphics[scale=0.5]{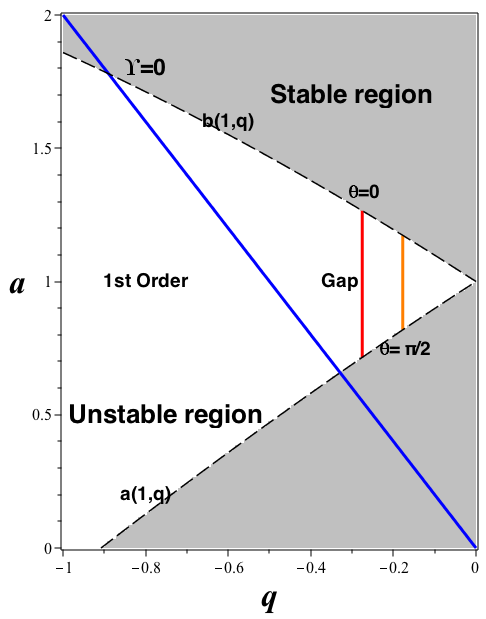}
 \caption{\label{Gap} Amplification of the energy spectrum for the case of long wavelength near of $\Upsilon=0$, i.e. $\epsilon=0$. For a  fixed field ($E_{0}=2$ V/m), the figure shows two energy first order gaps for $\theta=0$ and $\theta= \pi/2$, respectively; and two energy bands.We observe that the gap size decrease while $\theta$ increase for $0 \leq \theta \leq \pi/2$. }
 \end{figure}
As it is well documented \cite{book:598780}, the stability of Mathieu functions depends on the parameters $a$ and $q$. In Fig. \ref{Spectrum of Mathieu equations}, the white regions in the $(a,q)-$plane are those for which the solutions are unstable  and therefore, are not acceptable. On the other hand, the gray regions are those for which the solutions are acceptable wavefunctions. The boundaries between these regions are determined by the eigenvalues, $a_{n}(q)$ and $b_{n}(q)$, corresponding to the $2 \pi$- periodic Mathieu functions of integer order, $ce_{n}(q,\phi)$ and $se_{n}(q,\phi)$, respectively \cite{book:598780}. As a consequence of the Mathieu solutions mentioned above, a band structure naturally emerges in our problem. However, in the present case there is an extra constraint that relates $a$ and $q$ due to Equation \eqref{definition of a}. For fixed energies $\epsilon$, this condition is equivalent to drawing a set of parallel lines in Fig. \ref{Spectrum of Mathieu equations}. Since $\Upsilon^{2}$, where $\Upsilon=(\epsilon / \hbar \Omega)$, is always positive, in Fig. \ref{Spectrum of Mathieu equations} we plot the line $a=-2q$, which divides the spectrum into two zones. Energies on the left side of this line are strictly not allowed, while the values on the right side are allowed. States over this line have $\Upsilon=0$. Notice that for a fixed electric field and an angle $\theta$, $q$ is constant.Using this method, a simple analysis of Fig. \ref{Spectrum of Mathieu equations} reveals bands separated by energy gaps. The opening of these gaps is due to the space-time diffraction of electrons in phase with the electromagnetic field, and effect akin to the magnetoacoustic diffraction of electrons in phase with acoustic waves \cite{davidov1981teorÃ­a,Landau:1946jc}. One can think as if electrons in borophene acquire an effective mass under electromagnetic radiation, leading to a metal-insulator transition. This effect also occurs in graphene \cite{PhysRevB.78.201406,doi:10.1080/14786431003757794,PhysRevB.81.165433}.
 
Since this approach is for intense electric fields, let us estimate the size of these gaps. For a typical microwave frequency $\Omega= 50 \,\,\,GHz$ with an intensity  $E_{0}=2 \,\, V/m$ of the electric field, in Appendix B we show that the first order gap size is $\Delta \approx 0.024 \,\,meV$ for $\theta=0$.

\section{Conclusions \label{Conclusions}}
In conclusion, we have found an equation that describes the interaction between carriers in borophene under electromagnetic radiation. By solving this equation we found the energy spectrum for intense electric field (or long wavelength). The main features of the spectra is that bands appear and are separated by energy gaps; beside, the angle $\theta$ is important for the gap size. Thus, there is a metal-insulator transition and the conductivity can be controlled by applying an electromagnetic wave. Furthermore, our analysis reveals that electrons in borophene acquire an effective mass. This is due to the diffraction of electrons in phase with the electromagnetic wave, as happens in magnetoacoustic effects in metals. 

\section*{Acknowledgements}
We thank DGAPA-UNAM project IN-102717 for financial support. A.E. Champo thanks Instituto de F\'isica hospitality during a visit to finish this work. 

\appendix
\section{}
In this appendix, we derive Eqs. \eqref{Equation GammaA}, \eqref{Equation GammaB} and \eqref{Second Order Equation Spinor Chi 1}. We start from  Eqns. \eqref{Dirac Equation PsiA}, \eqref{Dirac Equation PsiB} and \eqref{Form of a solution} which are explicitly given by:
\begin{eqnarray} \label{Equation Phi A Notes}
\frac{d \Phi_{A}(\phi)}{d \phi}= \frac{i}{c} \lbrace \left[ v_{t} (\tilde{k}_{y}- \tilde{\xi} \cos \phi \sin \theta)- v_{y} \tilde{E} \right] \Phi_{A}(\phi)+ \nonumber \\
 \left[ v_{x} (\tilde{k}_{x}-\tilde{\xi} \cos \phi \cos \theta)-i v_{y}(\tilde{k}_{y}-\tilde{\xi}\cos \phi \sin \theta)\right] \Phi_{B}(\phi) \rbrace \nonumber \\
\,\, 
\end{eqnarray}
\begin{eqnarray} \label{Equation Phi B Notes}
\frac{d \Phi_{B}(\phi)}{d \phi}= \frac{i}{c} \lbrace \left[ v_{t} (\tilde{k}_{y}- \tilde{\xi} \cos \phi \sin \theta)- v_{y} \tilde{E} \right] \Phi_{B}(\phi)+ \nonumber \\
 \left[ v_{x} (\tilde{k}_{x}-\tilde{\xi} \cos \phi \cos \theta)+i v_{y}(\tilde{k}_{y}-\tilde{\xi}\cos \phi \sin \theta)\right] \Phi_{A}(\phi) \rbrace \nonumber \\
 \,\,
\end{eqnarray}
where $\tilde{E}= \frac{E}{\hbar v_{y} G}, \tilde{{\bf{k}}}= \frac{1}{G} {\bf{k}}, \tilde{\xi}=\frac{eE_{o}}{c \hbar G \Omega}$ and $c= \frac{\Omega}{G}$. Now, we propose to use the following transformation,
\begin{equation} \label{transformation PhiRho Notes}
\Gamma_{\rho} (\phi)=\exp{\lbrace-\frac{i}{c}[(v_{t}\tilde{k}_{y}-v_{y} \tilde{E})\phi-v_{t} \tilde{\xi} \sin \phi \sin \theta]\rbrace}\Phi_{\rho}(\phi)   
\end{equation}

Then, we reduce the Equations \eqref{Equation Phi A Notes} and \eqref{Equation Phi B Notes} to the following form:
\begin{equation} \label{Equation GammaA Notes}
\frac{d \Gamma_{A}(\phi)}{d \phi}= i C^{*}(\phi) \Gamma_{B}(\phi)
\end{equation}
\begin{equation} \label{Equation GammaB Notes}
\frac{d \Gamma_{B}(\phi)}{d \phi}= i C(\phi) \Gamma_{A}(\phi)
\end{equation}
where $C(\phi)=[v_{x}(\tilde{k}_{x}-\tilde{\xi}\cos \phi \cos \theta)+iv_{y}(\tilde{k}_{y}-\tilde{\xi}\cos \phi \sin \theta)]/c$.
 
From Eq. \eqref{Equation GammaA Notes}, \eqref{Equation GammaB Notes} we can obtain the following equations:
\begin{equation} \label{Second Order Equation GammaA Notes}
\frac{d^{2} \Gamma_{A}(\phi)}{d \phi^{2}}- \frac{1}{C^{*}(\phi)} \frac{d C^{*}(\phi)}{d \phi} \frac{d \Gamma_{A}(\phi)}{d \phi}+ |C(\phi)|^{2} \Gamma_{A}(\phi)=0
\end{equation}

\begin{equation} \label{Second Order Equation GammaB Notes}
\frac{d^{2} \Gamma_{B}(\phi)}{d \phi^{2}}- \frac{1}{C(\phi)} \frac{d C(\phi)}{d \phi} \frac{d \Gamma_{B}(\phi)}{d \phi}+ |C(\phi)|^{2} \Gamma_{B}(\phi)=0
\end{equation}

Inserting Eq. \eqref{Equation GammaA Notes} into Eq. \eqref{Second Order Equation GammaA Notes}, and Eq. \eqref{Equation GammaB Notes} into Eq. \eqref{Second Order Equation GammaB Notes}, we obtain that:
\begin{equation} \label{Second Order Equation GammaA 1 Notes}
\frac{d^{2} \Gamma_{A}(\phi)}{d \phi^{2}}- i \frac{d C^{*}(\phi)}{d \phi} \Gamma_{B}(\phi)+ |C(\phi)|^{2} \Gamma_{A}(\phi)=0
\end{equation}

\begin{equation} \label{Second Order Equation GammaB 1 Notes}
\frac{d^{2} \Gamma_{B}(\phi)}{d \phi^{2}}-i  \frac{d C(\phi)}{d \phi} \Gamma_{A}(\phi)+ |C(\phi)|^{2} \Gamma_{B}(\phi)=0
\end{equation}
And $$\frac{d C(\phi)}{d \phi}= \tilde{\xi} \sin \phi [(v_{x}/c) \cos \theta + i (v_{y}/c) \sin \theta]= \Lambda \tilde{\xi} \sin \phi e^{i \beta}$$ where $\Lambda= (1/c) \sqrt{v_{x}^{2}\cos^{2} \theta + v_{y}^{2} \sin^{2} \theta}$ and $\beta= \arctan((v_{y}/v_{x})\tan \theta)$.

We consider the transformation:
\begin{equation} \label{Second transformation GammaA Notes}
\Gamma_{A}(\phi)= e^{-i (\beta/2)}  \chi_{A} (\phi)
\end{equation}
\begin{equation} \label{Second transformation GammaB Notes}
\Gamma_{B}(\phi)= e^{i (\beta/2)}  \chi_{B} (\phi)
\end{equation}
besides, if we consider that:
\begin{equation} \label{Spinor Chi Notes}
{\bf{\chi}}(\phi)= \left(\begin{array}{lcc}
\chi_{A}(\phi) \\
\chi_{B} (\phi)
\end{array} \right)
\end{equation}
then, we can reduce equations \eqref{Second Order Equation GammaA 1 Notes}, \eqref{Second Order Equation GammaB 1 Notes} to the following Eq.
\begin{equation} \label{Second Order Equation Spinor Chi Notes}
\frac{d^{2}}{d \phi^{2}} {\bf{\chi}}(\phi)+ \left[ |C(\phi)|^{2}- i (\Lambda \tilde{\xi})\sigma_{x} \sin \phi  \right]{\bf{\chi}}(\phi)=0
\end{equation}
Now, observe that:
\begin{eqnarray}
\Lambda \tilde{\xi}= \frac{1}{\hbar \Omega} \zeta
\end{eqnarray}
where $\zeta= (eE_{0}/c \Omega) \sqrt{v_{x}^{2}\cos^{2} \theta + v_{y}^{2} \sin^{2} \theta}$, also:
\begin{equation}
|C(\phi)|^{2}= \left( \frac{1}{\hbar \Omega} \right)^{2} \left[ \epsilon^{2} + \zeta^{2} \cos^{2} \phi - 2 \zeta \frac{\vec{\nu} \cdot \vec{\kappa}}{|\vec{\nu}|} \cos \phi \right] 
\end{equation}

where $\epsilon= \hbar \sqrt{(v_{x} k_{x})^{2} +(v_{y}k_{y})^{2}}$, $\vec{\nu}=(v_{x} \cos \theta, v_{y} \sin \theta)$ and  $\vec{\kappa}= \hbar (v_{x} k_{x}, v_{y} k_{y})$. Therefore, Eq. \eqref{Second Order Equation Spinor Chi Notes} can be written as:
\begin{equation} \label{Second Order Equation Spinor Chi 1 Notes}
\frac{d^{2}}{d \phi^{2}} {\bf{\chi}}(\phi)+ F(\phi){\bf{\chi}}(\phi)=0
\end{equation}
with $F(\phi)$ defined as:
\begin{eqnarray} \label{Definition of F function Notes}
F(\phi)= \left( \frac{1}{\hbar \Omega} \right)^{2} \left[  \zeta^{2} \cos^{2} \phi - 2 \zeta \frac{\vec{\nu} \cdot \vec{\kappa}}{|\vec{\nu}|} \cos \phi \right] \nonumber \\ 
+\left( \frac{\epsilon}{\hbar \Omega} \right)^{2}- i\frac{\zeta}{\hbar \Omega}  \sigma_{x} \sin \phi  
\end{eqnarray}

Remark: If we define $v_{F}=v_{x}=v_{y},v_{t}=0, \theta=0$ then Eq. \eqref{Second Order Equation Spinor Chi 1 Notes} is reduced to the case of graphene as expected \cite{doi:10.1080/14786431003757794}.

\section{}
Let us estimate the gap size for a microwave frequency $\Omega= 50$ GHz with an intensity $E_{0}=2$ V/m of the electric field, first, we calculate the value of $q$ given by:
\begin{equation}
q(\theta)=-\left[ \frac{\sqrt{(v_{x}^{2} \cos \theta)^{2}+(v_{y} \sin \theta)^{2}} }{2\hbar \Omega} \left(\frac{e E_{0}}{c \Omega}\right) \right]^{2}
\end{equation}
Thus for $\theta=0$, then:
\begin{equation}
q(0)=-0.2754
\end{equation}
and for $\theta= \pi/2$, we obtain:
\begin{equation}
q(\pi/2)=-0.1773
\end{equation}
Thus, the gap of $n-th$ order is obtain by
\begin{equation}
\Delta_{n}(\theta)= \hbar \Omega \sqrt{|b_{n}(q)-a_{n}(q)|}
\end{equation}
Then, we evaluate $b_{n}(q)$ and $a_{n}(q)$ for the first order gap:
\begin{eqnarray}
\Delta_{1}(0) \approx 0.0243 \,\, meV\nonumber \\
\Delta_{1}(\pi/2) \approx  0.0195\,\,meV
\end{eqnarray}

\bibliographystyle{unsrt}
%\bibliography{biblio}

%\bibliography{McLachlan1947,Lopez-Rodriguez2010,Wen2018,Islam2017,Verma2017,Sadhukhan2017,Castro2009,Naumis2017,Nakhaee2018,Zabolotskiy2016,Lopez2016,Xu2016,Zhou2014,Wang2015,Balatsky2014,Mannix2015,Zhang2017}

\end{document}